\documentclass[12pt,a4paper]{article}
\usepackage[margin=2.2cm]{geometry}
\usepackage{authblk}

\usepackage[breaklinks]{hyperref}
\usepackage{booktabs}
\usepackage{mathptmx}

\usepackage[super,compress]{cite}

\usepackage{amsmath,amssymb,amscd,mathrsfs}
\usepackage{todonotes,soul}

\newcommand{\refcite}[1]{[\citen{#1}]}

\usepackage{amsthm}
\newtheorem{theorem}{Theorem}

\newcommand{\CoinX}[1]{C_0^\infty(#1)}
\newcommand{\ip}[2]{\langle #1\mid #2\rangle}

\DeclareMathOperator{\WF}{\textrm{WF}}
\DeclareMathOperator{\supp}{supp}
\DeclareMathOperator{\sinc}{sinc}
\DeclareMathOperator{\Sol}{Sol}

\newcommand{\id}{\textrm{id}}

\newcommand{\rce}{\textrm{rce}}

\newcommand{\dvol}{\textrm{dvol}} 

\newcommand{\FF}{{\mathscr{F}}}
\newcommand{\HH}{{\mathscr{H}}}

\newcommand{\Nc}{\mathcal{N}}
\newcommand{\Vc}{\mathcal{V}}
\newcommand{\Af}{\mathfrak{A}}
\newcommand{\Wf}{\mathfrak{W}}
\newcommand{\Mb}{{\boldsymbol{M}}}
\newcommand{\Nb}{{\boldsymbol{N}}}

\newcommand{\uk}{\mathbf{k}}
\newcommand{\ux}{\mathbf{x}}
\newcommand{\op}{\textrm{op}}
\newcommand{\SJ}{\textrm{SJ}}

\newcommand{\CC}{{\mathbb{C}}}
\newcommand{\RR}{{\mathbb{R}}}
\newcommand{\NN}{{\mathbb{N}}}

\begin{document}

\markboth{C.J. Fewster}
{The art of the state}

%
%

\title{The art of the state}

\author[1]{Christopher J. Fewster\thanks{\tt chris.fewster@york.ac.uk}}
\affil{Department of Mathematics,
	University of York, Heslington, York YO10 5DD, United Kingdom.}

\date{\today}
\maketitle
%
%
%
%

\begin{abstract}
	Quantum field theory on curved spacetimes lacks an obvious distinguished vacuum state. We review a recent no-go theorem that establishes the impossibility of finding a preferred state in each globally hyperbolic spacetime, subject to certain natural conditions. The result applies in particular to the free scalar field, but the proof is model-independent and therefore of wider applicability. In addition, we critically examine the recently proposed `SJ states', that are determined by the spacetime geometry alone, but which fail to be Hadamard in general. We describe a modified construction that can yield an infinite family of Hadamard states, and also explain recent results that motivate the Hadamard condition without direct reference to ultra-high energies or ultra-short distance structure.
\end{abstract}




\section{Introduction}

%
%

In quantum field theory (QFT) on Minkowski space, Poincar\'e invariance can be used to select a state of maximal symmetry and minimal energy (expressed by the spectrum condition) and under suitable conditions this provides a unique choice. This vacuum state has many
nice mathematical properties, which form the foundation for standard formulations
of QFT in flat spacetime. For example, 
translational invariance allows for the use of Fourier space, while
invariance under the Poincar\'e group 
allows for the identification of single-particle states via Wigner's analysis. 
From the physical viewpoint, it is important
that every inertial observer recognises the state as distinguished, for there
is no privileged inertial frame of reference in Minkowski space. 

These basic and important facts make it very tempting to seek a preferred state in QFT on curved spacetimes. The fundamental problem to be faced is that  
a generic curved spacetime has a trivial symmetry group. In such spacetimes, \emph{every} state is a state of maximal symmetry! Closely related to this is that
the notion of particle loses its sharp mathematical definition. At the physical level, there remains an operational definition of a particle in curved spacetime: namely, we could assert that a particle is present if a particle detector--an apparatus constructed using the blueprint for a particle detector that functions well in Minkowski space--is triggered. One expects that this would be provide a reasonable notion of `particle' provided that the spacetime curvature scales are much larger than the geometrical scales required for the construction and operation of the detector. However, the fact that detectors following different worldlines
respond differently to the same quantum state\cite{Unruh:1976,deBievreMerkli:2006} shows that this operational definition represents a retreat from the absolute notion of a particle available in the Minkowski theory. All in all, the absence
of a preferred state is often very awkward. As Afshordi et al remark in Ref.~\refcite{AAS}: 
\emph{It... ...seems unsatisfactory that as it stands, QFT lacks a general notion of ``vacuum'' which extends very far beyond flat spacetime}. 

On the other hand, one cannot wish away difficulties, and the contrasting view is that
one must simply learn to live without a preferred state. Wald has expressed this view as follows:
\emph{it is fruitless to seek a preferred vacuum state for general spacetimes...  ...the theory should be formulated in a manner that does not require one to specify a choice of state (or representation).}\cite{Wald:2009}
Instead, one should work with a class of physical states, such as the Hadamard class.

This paper serves two purposes. First, in Section~\ref{sec:nogo}, it will explain why (no matter how unsatisfactory this may seem) there can be no preferred state of quantum field theory that can be implemented in all spacetimes, and which obeys reasonable physical properties. This has long been part of the folk wisdom, illustrated by Wald's statement just quoted, but was proved rigorously only a few years ago.\cite{FewVer:dynloc_theory}
As will be described, this proof was
not only the first complete no-go theorem on preferred states for the free scalar field, it is formulated and proved for general quantum field theories in curved spacetimes (including the free scalar field as a special case).

Second, it will review the construction of `SJ states' that have been proposed recently for the free scalar field on curved spacetimes\cite{AAS}
(the analogue for Dirac fields\cite{FewsterLang:2015} has its roots in the `fermionic projector' formalism\cite{FinsterReintjes:2015} which pre-dates Ref.~\refcite{AAS}). These states can be shown to exist on a large class of spacetimes\cite{FewsterVerch-SJ} and depend only on the spacetime geometry and causal structure. However, they suffer from two problems\cite{FewsterVerch-SJ}: first, they depend on the geometry in a global fashion, which raises some operational questions (how would one prepare such a state, for example?). Second, at least in some concretely computable situations, they are not Hadamard. Despite these objections, SJ states have produced some interesting new insights for QFT in CST:
on one hand, the construction has been modified\cite{BruFre:2014,FewsterLang:2015} so as to produce new classes of Hadamard states in some spacetimes; on the other, it has prompted reflection on the nature of the Hadamard condition and how it can be motivated without direct recourse to the structure of the two-point function at short scales.\cite{FewsterVerch-NecHad} As these discussions involve the reformulation
of the Hadamard condition in terms of microlocal analysis,\cite{Radzikowski_ulocal1996,StVeWo:2002} the relevant concepts will be explained in Section~\ref{sec:prelim}.

\section{Preliminaries on quantum field theory in curved spacetimes}\label{sec:prelim}

\subsection{The algebraic approach} 

Before one can discuss which states are appropriate, it is necessary to 
describe the observables on which the states will be evaluated. The algebraic
approach to QFT in CST provides an ideal setting for these
purposes, because it cleanly separates the specification of observables and their algebraic relations from the discussion of states. By contrast, other treatments 
(e.g., as represented in Ref.~\refcite{BirrellDavies}) construct the Hilbert space and field operators together, thus presupposing a theory formulated on a Fock space and building in specific choices of positive or negative frequency modes from the start.
 
Consider a Lorentzian spacetime $M$ of dimension $n$, with metric $g$ of signature ${+}{-}\cdots{-}$. Assuming $(M,g)$ is time-oriented, we may introduce
the causal future $J^+(p)$ and past $J^-(p)$ of any point $p\in M$ as usual. The spacetime is \emph{globally hyperbolic} if there are no closed causal curves in $(M,g)$ and every set of the form $J^+(p)\cap J^-(q)$ is compact.\cite{Bernal:2006xf} This is equivalent to older definitions\cite{HawkingEllis} and implies that the spacetime can be foliated into smooth spacelike Cauchy surfaces.\cite{Bernal:2004gm,Bernal:2005qf} For brevity, we
use boldface $\Mb$, to denote the manifold, metric and time orientation of a globally hyperbolic spacetime. 

On globally hyperbolic spacetimes, the inhomogeneous Klein--Gordon equation 
\begin{equation}\label{eq:KGinhom}
P\phi:=(\Box+m^2+\xi R)\phi=f
\end{equation}
admits unique advanced and retarded Green operators $E^\pm$, mapping compactly supported test-functions $f$ to smooth solutions to \eqref{eq:KGinhom}, obeying
\begin{equation}
P E^\pm f = f, \qquad E^\pm P f = f, \qquad \supp E^\pm f\subset J^\pm(\supp f)
\end{equation}
for all test functions $f$, where $\supp$ denotes the support of a function -- the closure of the set of points at which it is nonzero. (See, e.g.,~\refcite{BarGinouxPfaffle} for the relevant theory of hyperbolic partial differential equations). The space of smooth solutions to the homogeneous Klein--Gordon equation that have compactly supported Cauchy data on some (and hence any) Cauchy surface will be denoted $\Sol(\Mb)$, and $\Sol(\Mb;\RR)$ 
for real-valued solutions. These spaces may be equipped with a symplectic form
\begin{equation}
\sigma(\phi_1,\phi_2) = \int_\Sigma \left(\phi_1 \nabla_n \phi_2 -\phi_2 \nabla_n \phi_1\right) \textrm{d}\Sigma ,
\end{equation}
where $\Sigma$ is any smooth spacelike Cauchy surface with future-pointing unit timelike normal $n^a$. Every
$\phi\in \Sol(\Mb)$ may be written as $\phi=E f$ for some compactly supported test function $f$ (which may be taken as real-valued if $\phi$ is, and may also be chosen to have support in an arbitrary neighbourhood of any Cauchy surface), where $E=E^{-}-E^{+}$ is the difference of the advanced and retarded Green operators. The symplectic form is then
\begin{equation}
\sigma(E f_1, E f_2) = E(f_1,f_2) := \int_M f_1(p) (Ef_2)(p)\, \dvol(p). 
\end{equation}

There are two main algebraic quantizations of this system (see Ref.~\refcite{BucGrun:2008} for another) in terms of complex algebras with a unit and antilinear $*$ involution. The first is the Weyl algebra $\Wf(\Mb)$, which is the unique (up to isomorphism) $C^*$-algebra specified by generators
$W_\Mb(\phi)$ labelled by $\phi\in \Sol(\Mb;\RR)$ and obeying the relations
\begin{equation}
W_\Mb(\phi)^*=W_\Mb(-\phi), \qquad 
W_\Mb(\phi_1)W_\Mb(\phi_2)=W_\Mb(\phi_1+\phi_2) e^{-i\sigma(\phi_1,\phi_2)/2}
\end{equation}
for all $\phi,\phi_1,\phi_2\in\Sol(\Mb;\RR)$. It follows that $W_\Mb(0)$ is the 
algebra unit, $W_\Mb(0)=\boldsymbol{1}$. 

On the other hand, the infinitesimal Weyl algebra $\Af(\Mb)$ has
generators $\Phi_\Mb(f)$ labelled by (complex-valued) test functions $f$ and subject
to the relations
\begin{equation}
\Phi_\Mb(f)^*=\Phi_\Mb(\overline{f}), \quad \Phi_\Mb(P_\Mb f)=0, \quad
[\Phi_\Mb(f_1),\Phi_\Mb(f_2)]= iE_\Mb(f_1,f_2)\boldsymbol{1},
\end{equation}
where the bar denotes a complex conjugation, and we additionally assume that
the labelling $f\mapsto \Phi_\Mb(f)$ is linear. This algebra can be made into
a topological $*$-algebra, by demanding continuity of the labelling in the test-function topology, but we will not do this here. The interpretation
is that $\Phi_\Mb(f)$ is a smeared quantum field, while $W_\Mb(\phi)$ are
exponentiated fields, formally related by 
\begin{equation}
W_\Mb(E_\Mb f) = e^{i\Phi_\Mb(f)}\qquad\text{(formal)}
\end{equation}
for real-valued $f$. This relation does not hold literally in the algebras ($\Af(\Mb)$ does not support a functional calculus, or the definition
of exponentials via power series) but does hold in suitably regular Hilbert space representations, in which the necessary analytic details can be resolved.

In either case, a state is a linear map $\omega$ from the algebra $\Wf(\Mb)$ or $\Af(\Mb)$ to the complex numbers which is normalized by $\omega(\boldsymbol{1})=1$  and positive in the sense that $\omega(A^*A)\ge 0$ for all $A$ in the algebra.
Physically, $\omega(A)$ is the expectation value of observable $A$ in the state $\omega$.  Using the GNS construction, every state determines (up to unitary equivalence) a Hilbert space representation
$\pi_\omega$ and a vector in the Hilbert space $\Omega_\omega$ so that
\begin{equation}\label{eq:GNS}
\omega(A) = \ip{\Omega_\omega}{\pi_\omega(A)\Omega_\omega}
\end{equation}
for all $A$ in the algebra and so that vectors of the form $\pi_\omega(A)\Omega_\omega$ are dense (that is, $\Omega_\omega$ is \emph{cyclic} for the algebra in the representation $\pi_\omega$). 
 
Not all states are physically relevant. For example, we may define a state $\omega_{\text{tr},\Mb}$ on $\Wf(\Mb)$ by
\[
\omega_{\text{tr},\Mb}(W(\phi)) =\begin{cases} 1 & \phi=0 \\ 0 & \text{otherwise},
\end{cases}
\]
and extending by linearity and continuity to all of $\Wf(\Mb)$. Thus, 
if $A$ and $B$ are convergent series 
\begin{equation}
A= \sum_{\phi\in\Sol(\Mb;\RR)} a_\phi W(\phi),\qquad B=\sum_{\phi\in\Sol(\Mb;\RR)} b_\phi W(\phi)
\end{equation}
where in each case at most countably many of the coefficients are nonzero, then
\begin{equation}
\omega_{\text{tr},\Mb}(AB) = \sum_{\phi\in\Sol(\Mb;\RR)} a_\phi b_{-\phi}
\end{equation}
as a consequence of the Weyl relations, and the series converges. Obviously, $\omega_{\text{tr},\Mb}(AB)=\omega_{\text{tr},\Mb}(BA)$; in other words,
$\omega_{\text{tr},\Mb}$ has the same properties as a matrix trace, for which reason it
is called the \emph{tracial state}. Mathematically, this state has its
uses: for example, the Weyl algebra can be constructed concretely on
the Hilbert space $\HH=\ell^2(\Sol(\Mb;\RR))$, the (inseparable) space
of square-summable sequences $a=(a_\phi)$ indexed by $\phi\in\Sol(\Mb;\RR)$, by
\begin{equation}
(W(\phi')a)_\phi = e^{-i\sigma(\phi',\phi)/2} a_{\phi+\phi'},
\end{equation}
whereupon the tracial state is realised by the vector $\Omega_{\text{tr},\Mb}=\delta_{\phi,0}$: i.e., $\omega_{\text{tr},\Mb}(A)=\ip{\Omega_{\text{tr},\Mb}}{A\Omega_{\text{tr},\Mb}}$. Physically, however, the tracial
state is pathological and is interpreted as having infinite temperature, as can be seen on recalling that KMS condition for inverse temperature $\beta$ requires $\omega_\beta(BA)=\omega_\beta(A\alpha_{i\beta}(B))$ where $\alpha_\tau$ is the automorphism of time-translation through $\tau$ (in cases where the solution space $\Sol(\Mb;\RR)$ admits such a symmetry).  

In consequence, the class of states should be restricted by physically motivated conditions. The traditional choice for the real scalar field has been to prefer the class of Hadamard states,
which we now briefly review. See Ref.~\refcite{KhavkineMoretti:2015} for a recent and more comprehensive pedagogical review.

\subsection{Hadamard states}\label{sec:Hadamard}

The Hadamard condition was originally formulated as a minimal deformation of the structure
of the Minkowski two-point function consistent with a curved background. Pragmatically, 
it is preferred because of its technical utility, and also because many specific states 
of interest turn out to be Hadamard, including ground and thermal states in 
stationary spacetimes (see, e.g., Ref.~\refcite{Sanders:2013} for general results). 

The Hadamard condition was first given a precise form by Kay and Wald.\cite{KayWald-PhysRep}
In essence, they define a state on $\Af(M)$ to be Hadamard if its two-point function
$W_2$ has the local form  
\begin{equation}\label{eq:Hadamard}
W_2(x,x')= 
\frac{U(x,x')}{4\pi^2\sigma_+(x,x')} + V(x,x')\log(\sigma_+/\ell^2) +\textrm{smooth},
\end{equation} 
where $U$ and $V$ are specified locally and geometrically, $\sigma(x,x')$ is the signed squared geodesic separation between $x$ and $x'$, $F(\sigma_+)$ denotes a particular
$i\epsilon$-regularisation of $F(\sigma)$ and $\ell$ is an arbitrary length scale.
The full definition is rather more complex: the geodesic separation is only defined when $x$ and $x'$ belong to a convex normal neighbourhood and the function $V$ is actually specified as a power series with recursively defined coefficients obtained as solutions to transport
equations; moreover, this series does not converge on general spacetimes (unless one cuts off successive terms on smaller and smaller regions\cite{Ho&Wa01}). 
In Ref.~\refcite{KayWald-PhysRep} it was also assumed that the two-point function has no nonlocal singularities and can be defined in the above way in a suitable neighbourhood of a Cauchy surface. However, 
a number of qualitative features can be read off: in particular (a) the definition of $U$ entails $U(x,x)=1$, so the leading singularity is fixed to agree with that of the Minkowski vacuum two-point function, and (b) any two Hadamard states have two-point functions that differ by a smooth function. The latter point is obvious
from \eqref{eq:Hadamard} at least locally, which suffices for the definition of
expectation values of Wick products by point-splitting, but can be extended to a global statement, as described later.

The last general point is that there are infinitely many different Hadamard states, distinguished by the smooth parts (in the above sense) of their two-point functions,
and there is no canonical choice. Locally, one could try to construct a smooth part by solving transport equations, but this requires the specification of the lowest order term in a series expansion in such a way that the resulting two-point function defines a state of the theory. In particular, we must ensure that $W_2(\overline{f},f) =\omega(\Phi(f)^*\Phi(f))\ge 0$ for all complex-valued test functions. Actually, this procedure is used in the construction of a local Hadamard parametrix, with the undetermined lowest order term set to zero\cite{Wald_qft}. However, the crucial difference is that the local Hadamard parametrix need not obey the positivity condition.

A completely different approach to the definition of Hadamard states was developed
by Radzik\-owski,\cite{Radzikowski_ulocal1996} based on the \emph{wavefront set} of a distribution, which we briefly explain. The underlying idea is based on the fact that the more
smooth a function or distribution is, the faster its Fourier transform decreases. 
For example, any smooth compactly supported test function has transform 
decaying faster than any inverse power, while a $\delta$-distribution has
a constant transform that does not decay in any direction. The
wavefront set localises this property in both position and momentum space. Suppose that $u$ is a distribution on $\RR^n$. If $f$
is any test function with compact support (that is, $f$ vanishes outside a bounded
set) then $fu$ is a compactly supported distribution and one may define
its Fourier transform by 
\begin{equation}\label{eq:fourier}
\widehat{fu}(k) = u(fe_k) = \int u(x) f(x) e^{ik\cdot x} \, \textrm{d}^n x,
\end{equation}
where $e_k(x)=e^{ik\cdot x}$ (we adopt a nonstandard convention for signs in the Fourier transform). Here we think of $k$ a row vector and $x$ as a column vector
so $k\cdot x$ is their natural pairing under matrix multiplication. 

In~\eqref{eq:fourier}, the multiplication by $f$ cuts off all singularities of $u$ outside the support of $f$,
providing a localisation in position space. By choosing $f$ to be more sharply localised around some particular point $x$, we focus more and more on the singularities of $u$ at $x$. Localisation in momentum space is achieved by analysing the directions near which $\widehat{fu}(k)$ decays rapidly. One defines $(x,k)\in \RR^n\times (\RR^n\setminus\{0\})$ to be a \emph{regular direction} for $u$ if there is a compactly supported test function $f$ with $f(x)\neq 0$ and an open cone $V$ in $\RR^n\setminus\{0\}$ containing $k$ such that
\begin{equation}\label{eq:reg_dirn}
|\ell|^N \widehat{fu}(\ell) \longrightarrow 0 \qquad \textrm{as $\ell\to\infty$ in $V$,
for each $N\in\NN$.}
\end{equation}

The wavefront set $\WF(u)$ consists of all $(x,k)\in \RR^n\times (\RR^n\setminus\{0\})$ 
that are \emph{not} regular directions for $u$. A very simple example is given by the $\delta$-function, for which $\widehat{f\delta}(k)=f(0)$ for all $k$ and $f$. Thus
if $f(0)\neq 0$, the transform $\widehat{f\delta}$ does not decay in any direction, while if $f(0)=0$ it decays in all directions (in a rather trivial sense). From this, it is easily seen that 
\begin{equation}
\WF(\delta)=\{(0,k): k\in\RR^n\setminus\{0\}\}.
\end{equation}
A remarkable fact is that the wavefront set can also be defined for distributions on a general smooth manifold $X$. The Fourier transform has to be defined using local coordinates, but the
wavefront set transforms under change of coordinates as a subset of the cotangent bundle $T^*X$. As we still exclude zero covectors, we write $\dot{T}^*X=\{(x,k)\in T^*X: k\neq 0\}$
and so in fact $\WF(u)\subset \dot{T}^*X$.  

Radzikowski showed how the Hadamard condition could be expressed
in terms of the wavefront set $\WF(W_2)$, which is a subset of $\dot{T}^*(M\times M)$. 
Namely, he showed that if $W_2$ is the two-point function of a state on $\Af(M)$, then the state is Hadamard if and only if 
\begin{equation}\label{eq:uSC_Rad}
\WF(W_2)=\{(x,k;x',k')\in \dot{T}^*M\times \dot{T}^*M: (x,k)\sim (x',-k'),~k\in\Nc^+_x\},
\end{equation}   
where $\Nc_x^{+/-}$ are the cones of positive/negative-frequency null covectors at $x$,
and the relation $(x,k)\sim (y,\ell)$ means that there exists a null geodesic from $x$ to $y$,
with tangent vector $k^a$ at $x$, and $\ell^a$ at $y$, and with $\ell$ obtained by 
parallel propagation of $k$ along the geodesic. In the case $x=y$, we interpret this
to mean simply that $k=\ell$ is null. Furthermore, the additional global conditions
imposed by Kay and Wald\cite{KayWald-PhysRep} turn out to be unnecessary: it is enough that every point has a neighbourhood in which the two-point function takes the local Hadamard form~\eqref{eq:Hadamard}.\cite{Radzikowski_locglob}

Once one has understood the definition of the wavefront set, \eqref{eq:uSC_Rad} is a much simpler formulation of the Hadamard condition than Eq.~\eqref{eq:Hadamard}, especially recalling that our discussion of~\eqref{eq:Hadamard} suppressed considerable detail. Furthermore, it makes available many techniques
of microlocal analysis that allow for the efficient manipulation of distributions.
For example, wavefront sets can be used to express criteria under which distributions can be multiplied, restricted to submanifolds or pulled back by smooth maps of manifolds. Over the past 25 years, this form of the Hadamard condition has found many applications, particularly in the perturbative construction of interacting QFTs in curved spacetimes,\cite{BruFreKoe,BrFr2000,Ho&Wa01,Ho&Wa02} but also including the proof of mathematically rigorous
Quantum Energy Inequalities\cite{Fews00,FewsterVerch-DiracQWEI,Few&Pfen03},
the solution of semiclassical Einstein equations\cite{Pinamonti:2011},
the description of Unruh-deWitt detectors along general worldlines in curved spacetimes,\cite{JunkerSchrohe,LoukoSatz:2008,FewsterJAubryLouko:2016} and
the analysis of QFT on spacetimes with closed timelike curves.\cite{KayRadWald:1997}

Radzikowski's work built on the pioneering results of Duistermaat and H\"ormander on 
distinguished parametrices.\cite{DuiHoer_FIOii:1972} It is worth noting that Duistermaat and H\"ormander were well aware of potential applications in quantum field theory. Indeed, they showed that the Klein--Gordon equation (on suitable manifolds) has a parametrix analogous to the Feynman propagator, with an associated Wightman two-point distribution whose wave-front set is given by Eq.~\eqref{eq:uSC_Rad}. Further, they knew that
this distribution could be made positive by the addition of a suitable smooth part. 
However, their constructions were unique only up to the addition of further smooth corrections and Duistermaat and H\"ormander could see no way of
choosing one of them: as they put it, \emph{we do not see how to fix the indetermination}.\cite{DuiHoer_FIOii:1972} It is tempting to speculate on how QFT in CST might have developed, had Duistermaat and H\"ormander been searching for a class of two-point functions, rather than a distinguished one. As it was, it took 20 years before Radzikowski finally made the right connections.

A further simplification of the Hadamard condition was made by Strohmaier, Verch and Wollenberg.\cite{StVeWo:2002}
Consider any unit vector $\Omega$ in a representation of $\Af(M)$ on a Hilbert space $\HH$
so that $\Omega$ belongs to the operator domain of every smeared field operators $\Phi(f)$ 
in the representation. Then we may define a vector-valued distribution mapping test functions to elements of $\HH$, $f\mapsto \Phi(f)\Omega\in\HH$. This too, has
a wavefront set (defined exactly as above, requiring the limits in Eq.~\eqref{eq:reg_dirn} to exist in the sense of norm convergence on $\HH$). It was shown in Ref.~\refcite{StVeWo:2002} that $\Omega$ defines a Hadamard state if and only if
\begin{equation}\label{eq:uSC}
\WF(\Phi(\cdot)\Omega) \subset \Vc^-,
\end{equation}
where $\Vc^-\subset \dot{T}^*M$ is the bundle whose fibre at $x$ is the cone of all negative-frequency covectors at $x$. If Eq.~\eqref{eq:uSC} holds then one may deduce the stricter condition
\begin{equation}\label{eq:uSC2}
\WF(\Phi(\cdot)\Omega) = \Nc^-.
\end{equation}
This is a remarkable simplification relative to Eq.~\eqref{eq:Hadamard} and Eq.~\eqref{eq:uSC_Rad} and makes clear that the information conveyed by the
Hadamard condition is precisely that all the singularities in $\Phi(x)\Omega$ are negative-frequency.  Although we have only
discussed $2$-point functions, the condition~\eqref{eq:uSC2} has implications for all $n$-point functions: in particular they must obey the microlocal spectrum condition.\cite{Sanders:2010}. Reducing the
degree of regularity required one arrives at adiabatic states, which can also
be studied using microlocal techniques.\cite{JunkerSchrohe}

\section{A no-go theorem for natural states}\label{sec:nogo}

In this section, we describe an operationally motivated definition of a preferred state and then prove that (with suitable additional assumptions) such states cannot exist.

\subsection{Natural states}

States may be regarded as preparations for experimental measurements. In turn,  those measurements are conducted using pieces of apparatus described by a blueprint or instruction manual. An observer can use their reference frame of rods and clocks to construct and operate a piece of apparatus according to the blueprint. 

In Minkowski spacetime, the Poincar\'e invariance of the vacuum state  
means that observables $A$ and $A'$, constructed from the same blueprint but with respect to different inertial reference frames, will produce statistically equivalent measurement outcomes in the vacuum state: all realisations of the blueprint in inertial frames are equivalent as far as the results are concerned. 

Now consider measurements in curved spacetimes. An attractive idea is to replicate
the previous discussion as far as possible. Thus we require that equivalent realisations of the same experiment, undertaken in the preferred state, should yield the same measurement outcome statistics. Given that
experiments are conducted locally, in ignorance of the spacetime geometry beyond
the experimental region, this principle should be extended to equivalent experiments conducted in different global spacetimes. Therefore, if $\omega$ and $\omega'$ are the preferred states on spacetimes $\Mb$ and $\Mb'$ respectively, 
and $A$ and $A'$ represent identical experiments in these two spacetimes, one requires that the measurement statistics for $A$ in $\omega$ should coincide with those for $A'$ in $\omega'$. In particular, the expectation values should coincide:
$\omega(A)=\omega'(A')$. 

A special case, in which there is a clear meaning to `the same experiment'
in different spacetimes, arises when a spacetime $\Mb$ can be isometrically embedded within another, $\Nb$, in such a way that orientations and causal structure are preserved. Then any piece of apparatus constructed in $\Mb$ has a direct equivalent
in $\Nb$ and there are no causal connections between parts of the 
apparatus during the operation of the experiment in $\Nb$ that are not already present in the version of the experiment in $\Mb$. 

Being more formal, if $\psi:\Mb\to\Nb$ is an isometric, time- and space-orientation preserving embedding with a causally convex image one expects there to be a map
$\Af(\psi):\Af(\Mb)\to\Af(\Nb)$ between the respective observable algebras. 
It is natural to require that the map is a $*$-homomorphism that maps the unit
of $\Af(\Mb)$ to the unit in $\Af(\Nb)$. If every observable in $\Mb$ has an equivalent in $\Nb$, which would be expected for local observables (here, we 
exclude any observables sensitive to global spacetime topology) then $\Af(\psi)$
is injective. It is further natural to assume that the identity embedding should correspond to the identity mapping on algebras and that successive embeddings should behave in an obvious way:
\begin{equation}
\Af(\id_\Mb)=\id_{\Af(\Mb)} \qquad \Af(\psi\circ\varphi) = \Af(\psi)\circ\Af(\varphi).
\end{equation}
This line of thought has essentially brought us to the framework of locally
covariant quantum field theory,\cite{BrFrVe03} in which a theory is described as a functor from the category of globally hyperbolic spacetimes to a category of $*$-algebras. An extensive discussion of this viewpoint on QFT in curved spacetimes
can be found in Ref.~\refcite{FewVerchaqftincst:2015}, however our discussion
here can be made without recourse to category theory. In particular, the theory of the free scalar field in the infinitesimal Weyl algebra quantization is a locally covariant theory, with the maps 
$\Af(\psi)$ defined so that
\begin{equation}
\Af(\psi)\Phi_\Mb(f) =\Phi_\Nb(\psi_*f) ,
\end{equation}
where $\psi_*f$ is the push-forward of the test function $f$ under $\psi$. 
Similarly, the Weyl algebraic quantization is also locally covariant, with
$\Wf(\psi)W_\Mb(E_\Mb f)=W_\Nb(E_\Nb\psi_*f)$. 

We can now give a precise notion of a preferred state for
a locally covariant theory $\Af$, along the lines of our general discussion above.
Namely a \emph{natural state} for $\Af$ is a choice of state in each spacetime $\Mb$ so that  
\begin{equation}\label{eq:natstate}
\omega_\Mb(A)  = \omega_\Nb(\Af(\psi)A) \qquad (A\in\Af(\Mb))
\end{equation} 
holds whenever $\psi:\Mb\to\Nb$ is an allowed embedding of spacetimes. 
Written more abstractly, 
\begin{equation}\label{eq:natstate_abs}
\omega_\Mb = \omega_\Nb\circ\Af(\psi) = \Af(\psi)^*\omega_\Nb.
\end{equation}
Recalling that $\Af(\psi)A$ represents the same experiment in $\Nb$ as $A$ does in $\Mb$, this precisely requires that equivalent experiments should return equivalent measurement outcomes when performed in the preferred state.  

Note that any natural state exhibits maximal symmetry in each spacetime. For if $\psi:\Mb\to\Mb$ is an orientation  and time-orientation preserving isometric isomorphism, then $\Af(\psi)$ is an automorphism of $\Af(\Mb)$ and \eqref{eq:natstate_abs} asserts that $\omega_\Mb = \omega_\Mb\circ\Af(\psi)$. 

We now turn to the analysis of our definition and, ultimately, the no-go theorem.

\subsection{Natural states and relative Cauchy evolution}

\begin{figure}[pt]
	\centerline{\includegraphics[width=0.5\textwidth]{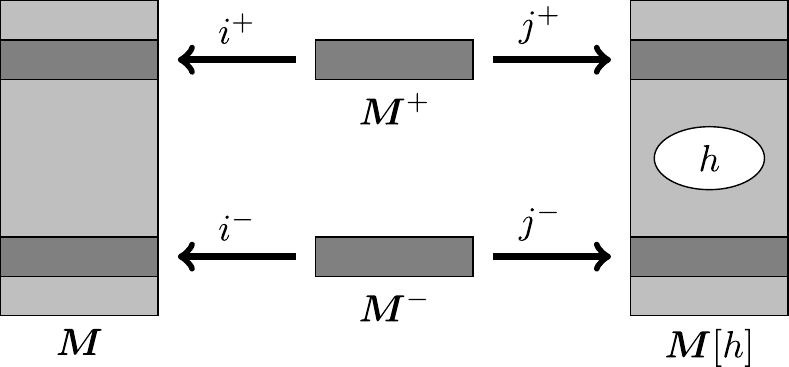}}
	\vspace*{8pt}
	\caption{Illustration of relative Cauchy evolution. All spacetimes shown are globally hyperbolic and all the maps are allowed embeddings whose images contain Cauchy surfaces. \label{fig:rce}}
\end{figure}

Consider~\eqref{eq:natstate_abs} in the situation where 
$\Af(\psi)$ is invertible, which occurs in particular when the image of $\psi$ contains a Cauchy surface of $\Nb$, by the timeslice property of locally covariant fields.\cite{BrFrVe03} In this case one has $\omega_\Nb =\omega_\Mb\circ\Af(\psi)^{-1}$.   
Now consider the situation illustrated in Fig.~\ref{fig:rce}.
Here we see a 
globally hyperbolic spacetime $\Mb$ and another, $\Mb[h]$, with the same underlying manifold as $\Mb$ but whose metric differs by the addition of a compactly supported metric perturbation $h$.
To the future of the perturbation, $\Mb$ and $\Mb[h]$ have common Cauchy surfaces
and a causally convex neighbourhood of one such has been chosen. This can be regarded as a spacetime $\Mb^+$ in its own right, embedded into $\Mb$ and $\Mb[h]$ by the maps $i^+$ and $j^+$ respectively (as functions between manifolds, they are identical inclusion maps). The spacetime $\Mb^-$ and embeddings $i^-$, $j^-$ are constructed in the same way, but to the past of the perturbation.

As all of the embeddings induce invertible maps of the corresponding algebras,
we have $\omega_{\Mb^+}=\omega_{\Mb}\circ \Af(i^+)$, $\omega_{\Mb[h]}=\omega_{\Mb^+}\circ \Af(j^+)^{-1}$ and so on, leading to the conclusion that 
\begin{equation}\label{eq:rceinvomega}
\omega_{\Mb}\circ \rce_{\Mb}[h] = \omega_{\Mb},
\end{equation} 
for all allowed metric perturbations $h$, where the \emph{relative Cauchy evolution}\cite{BrFrVe03} is defined by
\begin{equation}
\rce_{\Mb}[h] = \Af(i^-)^{-1}\circ\Af(j^-)\circ\Af(j^+)^{-1}\circ\Af(i^+)
\end{equation}
and is an automorphism of the algebra $\Af(\Mb)$ that encodes the response
of the theory to a metric perturbation. The relative Cauchy evolution is independent of the choices of $\Mb^\pm$ made, provided that they lie to the future and past of the perturbation $h$. It plays an important role in many applications of locally covariant theories -- in particular, the functional derivative with respect to the metric is related to the stress-energy tensor of the theory.\cite{BrFrVe03} One property of the relative Cauchy evolution that will be needed is that it is local: if $A\in\Af(\Mb)$ is localised in 
a region $R$ then $A$ is invariant under $\rce_{\Mb}[h]$ for every
metric perturbation $h$ supported in the causal complement of $R$.\cite{FewVer:dynloc_theory}

Equation~\eqref{eq:rceinvomega} seems a very strong constraint and 
might already lead one to conclude that no natural state can exist. But this conclusion
would be premature. Consider the locally covariant theory of the free scalar field, expressed
using the Weyl algebras. This theory admits a locally covariant natural state given by the trace state on each $\Wf(\Mb)$.\footnote{I thank Klaus Fredenhagen for
	this observation.} For let $\psi:\Mb\to\Nb$ and $\phi\in\Sol(\Mb;\RR)$. Then for any 
compactly supported real test function $f$ on $\Mb$,
\begin{equation}
(\Wf(\psi)^*\omega_{\text{tr},\Nb})(W_\Mb(E_\Mb f)) = \omega_{\text{tr},\Nb}(W_\Nb(E_\Nb\psi_* f))= \begin{cases} 1 & E_\Nb\psi_* f=0 \\
0 & \text{otherwise} \end{cases}
\end{equation}
and as $E_\Nb\psi_* f=0$ if and only if $E_\Mb f =0$ we see that $\Wf(\psi)^*\omega_{\text{tr},\Nb}=\omega_{\text{tr},\Mb}$ is the trace state on $\Wf(\Mb)$. To exclude such pathological examples, it is necessary to set out further conditions that should be required of a physical state.  This can be done for a general class of locally covariant theories and leads to a model-independent no-go result that we now describe. 
 
\subsection{The no-go result}

The no-go theorem proved by Verch and the present author\cite{FewVer:dynloc_theory} relies on a number of general assumptions, all of which were developed for other purposes in algebraic and locally covariant quantum field theory. The assumptions are met by standard models of the free scalar field, for example, but the proof is model-independent so our result has much wider applicability. The version stated here is not the most general.
\begin{theorem}\label{thm:nogo}
If a locally covariant theory $\Af$ admits a natural state and
\begin{itemize}
	\item the theory in Minkowski space $\Mb_0$ obeys standard AQFT assumptions
	with $\omega_{\Mb_0}$ as the Mink\-owski vacuum: in particular, (a) the GNS Hilbert space
	representation of $\Af(\Mb_0)$ induced by $\omega_{\Mb_0}$ is faithful, (b) subalgebras corresponding to spacelike separated subregions commute, (c) the GNS vector has the \emph{Reeh--Schlieder} property that it is cyclic for the subalgebra of any bounded subregion
	\item $\Af$ obeys \emph{dynamical locality} and \emph{extended locality}
\end{itemize}
then $\Af$ is trivial: each  $\Af(\Mb)$ consists only of multiples of the unit.
\end{theorem}
It would take us too far afield to give a precise definition of dynamical locality,
and we only sketch the idea below; see Ref.~\refcite{Fewster:2015} for a summary.  It is worth emphasising that dynamical locality was introduced for a different purpose, in a discussion of when a locally covariant theory can be said to represent the same physics in all spacetimes (abbreviated SPASs).\cite{FewVer:dynloc_theory} The no-go theorem was a spin-off result from this analysis. Extended locality\cite{Landau1969} is a much older idea and states that the subalgebras of
spacelike separated regions can intersect only in multiples of the unit. 
The proof of Theorem~\ref{thm:nogo} involves several steps and proceeds (in outline) as follows.

\paragraph{\emph{(a) Unitary implementation of Relative Cauchy evolution}}
A consequence of \eqref{eq:rceinvomega} is that the relative Cauchy evolution is unitarily implemented in the GNS Hilbert space representation induced by $\omega_\Mb$, provided that this representation is faithful. Specifically, we define an invertible linear map $U_\Mb$ on the dense subspace of vectors $\pi_\Mb(A)\Omega_\Mb$ by
\begin{equation}\label{eq:rce_implem}
U_\Mb[h]\pi_\Mb(A)\Omega_\Mb := \pi_\Mb(\rce_{\Mb}[h]A)\Omega_\Mb,
\end{equation}
which is well-defined by faithfulness of the GNS representation, and note that 
\begin{align}
\|U_\Mb[h]\pi_\Mb(A)\Omega_\Mb\|^2 &= \|\pi_\Mb(\rce_{\Mb}[h]A)\Omega_\Mb\|^2 = 
\omega_\Mb(\rce_{\Mb}[h] (A^*A)) \notag \\ &=\omega_\Mb(A^*A) = \|\pi_\Mb(A)\Omega_\Mb\|^2
\end{align}
for any $A$. Therefore $U_\Mb[h]$ is a bounded invertible isometry on a dense subspace and extends
to a unitary operator on the Hilbert space. Next, the calculation
\begin{align}
U_\Mb[h] \pi_\Mb(A) \pi_\Mb(B)\Omega_\Mb &= \pi_\Mb(\rce_{\Mb}[h] (AB))\Omega_\Mb 
\notag\\
&=  \pi_\Mb(\rce_{\Mb}[h] A)U_\Mb[h] \pi_\Mb(B)\Omega_\Mb
\end{align}
shows that 
\begin{equation}
U_\Mb[h] \pi_\Mb(A) U_\Mb[h]^{-1}=\pi_\Mb(\rce_{\Mb}[h] A) ,
\end{equation}
which is the desired unitary implementation property. Note also that
$U_\Mb[h]\Omega_\Mb = \Omega_\Mb$ by~\eqref{eq:rce_implem} with $A=\boldsymbol{1}$.

\paragraph{\emph{(b) Relative Cauchy evolution is trivial in Minkowski spacetime}}
Now specialise to Minkowski spacetime. We will show that $U_{\Mb_0}[h]$ and $\rce_{\Mb_0}[h]$ are trivial for every compactly supported metric perturbation $h$. For we may certainly find a spacetime region $O$ so that
$h$ is supported in $O^\perp$. By the locality property of the relative Cauchy evolution, $U_{\Mb_0}[h]$ leaves invariant all vectors $\pi_{\Mb_0}(A)\Omega_{\Mb_0}$ for $A$ localised in $O$. As these vectors form a dense set
due to the Reeh--Schlieder property, each $U_{\Mb_0}[h]$ is the identity operator,
so $\pi_\Mb(\rce_{\Mb}[h] A)=\pi_\Mb(A)$ for all $A$. 
Because the representation is faithful, this shows that $\rce_{\Mb_0}[h]$ is the identity automorphism for every compactly supported metric perturbation $h$. 

\paragraph{\emph{(c) The theory is trivial in Minkowski spacetime}}
The next step uses the dynamical locality and extended locality of $\Af$ and aims to show that $\Af(\Mb_0)$ is trivial, consisting only of scalar multiples of the unit. Simplifying somewhat, dynamical locality asserts that an  element $A\in\Af(\Mb_0)$ is localised in a region $O$ if and only if $\rce_{\Mb_0}[h]A=A$ for all metric perturbations supported in the causal complement of $O$. But as all relative Cauchy evolutions in $\Mb_0$ are trivial,
every $A$ can be localised in any region. Indeed, any $A$ can be localised in regions that are causally disjoint! Under the assumption of extended locality, this entails that every element is a multiple of the unit.

\paragraph{\emph{(d) The theory is trivial in all spacetimes}}
The arguments advanced so far show that $\Af(\Mb_0)$ is trivial. But what about
spacetimes that do not meet the conditions placed on $\Mb_0$? The \emph{coup de grace} is provided by a consequence of dynamical locality called the SPASs property (standing for `Same Physics in All Spacetimes'). This says (again suppressing technical detail) that if two locally covariant theories coincide in one spacetime, and one is a subtheory of the other, then
they are equivalent theories. In our case, the trivial locally covariant theory, whose algebras are just multiples of the unit in every spacetime, is a subtheory
of every other locally covariant theory. In particular, it is a subtheory of $\Af$, 
with which it coincides in $\Mb_0$ as shown above. Therefore $\Af$ is equivalent to the trivial theory, so every $\Af(\Mb)$ simply consists of multiples of the unit. 
This completes the proof of Theorem~\ref{thm:nogo}.

\subsection{Beyond the no-go result}

Theorem~\ref{thm:nogo} shows that natural states do not exist for nontrivial
locally covariant theories obeying certain additional hypotheses. 
We re-emphasise that this excludes the choice of a \emph{single} state for each spacetime.
The Hadamard states of the free scalar field provide a very good example of a 
\emph{class} of states that respects spacetime embeddings: for if a state $\omega$
is Hadamard on spacetime $\Nb$ and $\psi:\Mb\to\Nb$ is an allowed embedding, then
$\Af(\psi)^*\omega$ is a Hadamard state on $\Mb$; this is conveniently
seen using the microlocal form of the condition in~\eqref{eq:uSC_Rad}.\cite{BrFrVe03} The most 
obvious message of Theorem~\ref{thm:nogo} is to abandon the search for single distinguished states and instead seek distinguished classes of states.

However, a no-go theorem is only as strong as its hypotheses. What if 
one or more of these is dropped? Examination of the proof shows that the least familiar conditions -- dynamical locality and extended locality -- are needed only in steps (c) and (d). Even without them, one can still reach the conclusion that
the relative Cauchy evolution is trivial in Minkowski space, the infinitesimal version of which is that the stress-energy tensor commutes with all observables. This already rules out most interesting theories and might be taken as a sufficient, though weaker, no-go result. The other conditions on the Minkowski spacetime 
theory are sufficiently standard within algebraic QFT (and satisfied by models of interest) that there seems little mileage in dropping them. What we stated
as the Reeh--Schlieder property, for example, is part of the Reeh--Schlieder theorem, which follows from basic tenets of QFT in Minkowski spacetime.\cite{StreaterWightman,Haag} As we have also seen, some physical conditions are required to rule out pathological natural states such as the tracial state on the Weyl algebraic theory.
 
Therefore the only viable way to circumvent the no-go theorem is to reject or weaken the definition of a natural state as characteristic of the desired preferred state. 
Rejecting the definition would mean to reject the operational arguments that led to it, which raises its own problem: if the preferred state is not determined by the local geometry, then how can it be prepared (at least approximately) by an experimenter who can only access and influence the local geometry and quantum field in a local region? In that case, any preferred state that did exist would be (at best) of mathematical rather than physical interest.

A more promising direction is to drop the requirement that the preferred state exist in all globally hyperbolic spacetimes, and to define states with certain common features on a subclass.  The most familiar examples are provided by stationary spacetimes, where one can distinguish ground states and thermal states by general definitions. In the case of the scalar field, for example, these states also
have other good, physical properties; in particular they are Hadamard (see Ref.~\refcite{Sanders:2013} for detailed results to this effect). Other classes that have been investigated include asymptotically flat spacetimes, in which one can seek states invariant under the BMS group,\cite{Moretti:2008,DapMorPin:2009,DappSiemssen:2013} or Lorentzian scattering spacetimes, in which distinguished propagators and states can be constructed, and shown to be Hadamard -- see Refs.~\citen{VasyWrochna:2015,GerardWrochna:2017}
and references therein for details. 

One should raise a note of caution about the physical utility of mathematically distinguished states. As repeatedly mentioned, experiments are conducted in local regions of spacetime, without access to the limiting asymptotic regions. Therefore, without good information about the local properties of asymptotically-determined states, it is unclear how they could be
prepared or recognised in experiments. (It is also true that the asymptotic states sometimes depend on specific coordinates used in the construction, so they are not always uniquely distinguished by the geometry alone.) One might think that a similar objection
would attach to ground states. It is true that they invoke the global structure
of spacetime. However, it is possible to give an operational procedure for
creating a local approximation to the ground state: simply wait for any radiation
to leave the region, prevent any more from entering, and remove any bound state excitations. This is exactly what happens when a vacuum tube is prepared. In this
way one can create as large a region as desired in which the state is close to vacuum. It is important to supply similar operational understandings for other mathematically distinguished states.

\section{SJ states}\label{sec:SJ}

In 2012, Afshordi, Aslanbeigi \& Sorkin\cite{AAS} introduced a prescription that --- at least formally --- obtains a mathematically distinguished `SJ state' for the scalar field from the spacetime geometry alone. The initials SJ are used because the
proposal derives from work on causal set QFT due to Sorkin and Johnston.\cite{Johnston:2009,Sorkin:2011} In this section, we review and critically evaluate the proposal following Refs.~\citen{FewsterVerch-SJ,FewsterVerch-NecHad}, before explaining how the construction can be modified in some circumstances, so as to produce a class of Hadamard states.\cite{BrumFredenhagen:2014}

\subsection{The SJ proposal}

The construction can be described as follows. Let $\Mb$ be a globally hyperbolic spacetime and write $\HH$ for the Hilbert space $L^2(\Mb)$ of square-integrable functions with respect to the volume measure induced by the metric. The key idea is to regard the advanced-minus-retarded solution operator $E:\CoinX{M}\to C^\infty(M)$ as an operator on $\HH$. Clearly there are
immediate questions as to the square-integrability of functions $Ef$, where $f$ is a test function. However, let us set these to the side for the moment and proceed
formally, assuming that $E$ extends to an operator $E_\op$ on $\HH$.
The matrix elements of $E_\op$ reproduce the bi-distributional values of $E$:
\begin{equation}
\ip{\overline{f}}{E_\op g} = E(f,g) = \int_M f Eg\, \dvol_\Mb.
\end{equation}  
We use two properties of $E$. First, it is real,
in the sense that $\overline{E(f,g)} = E(\overline{f},\overline{g})$. This implies
$\Gamma E_\op \Gamma = E_\op$, where we write $\Gamma$ for the antilinear operation of complex conjugation. Second, $E$ is antisymmetric in its arguments, which gives
\begin{equation}
\ip{\overline{f}}{E_\op g} = -E(g,f)= -\ip{\Gamma g}{E_\op f} 
= -\ip{\Gamma E_\op f}{g} = -\ip{E_\op \overline{f}}{g}.
\end{equation}
So $E_\op$ is anti-self-adjoint, which means that $A=iE$ is self-adjoint, $A=A^*$.

The idea is now to decompose $A=A_++A_-$, where $A_+$ is a positive and $A_-$ a negative operator. Functional calculus on $\HH$ provides a distinguished decomposition, namely,
\begin{equation}\label{eq:Apm}
A_\pm = \frac{1}{2}\left(A\pm \sqrt{A^*A}\right) = \frac{1}{2}\left(A\pm \sqrt{A^2}\right),
\end{equation}
where the square root sign denotes the unique positive operator square root.
One can either accept \eqref{eq:Apm} as the basis of the SJ prescription, or
attempt to justify it in some way. 
Sorkin\cite{SorkinGF:2017} has suggested three requirements that could be taken as a more fundamental starting-point for the selection of a positive operator $A_\SJ$ from a given bounded self-adjoint operator $A$ (the labelling is ours):
\begin{description}
	\item[SJ1] $A_\SJ - \Gamma A_\SJ \Gamma = A$  
	\item[SJ2] $A_\SJ\Gamma A_\SJ \Gamma = 0$
	\item[SJ3] $A_\SJ\ge 0$
\end{description}
and has shown that if $A$ is a bounded Hilbert space operator, the unique solution is $A_\SJ = A_+$ as defined by \eqref{eq:Apm}. Actually, SJ1 and SJ3 are simply expressions of the CCRs and positivity of states so the main content of the SJ proposal is encapsulated in SJ2, which Sorkin calls the `ground state condition'.

With the positive part $A_+$ fixed,
the \emph{SJ two-point function} is now defined by
\begin{equation}
W_\SJ(f,g):=\ip{\Gamma f}{A_+ g}
\end{equation}
for all test functions $f,g$. As we now show, this has all the properties required
of the two-point function of a quasifree state on the algebra $\Af(\Mb)$.

The most obvious is positivity: by construction, 
\begin{equation}
W_\SJ(\overline{f},f)=\ip{f}{A_+f}\ge 0
\end{equation} 
for all test functions $f$.
Next, it is true of any positive operator $T$ that $\ker\sqrt{T}=\ker T$,\footnote{If $\sqrt{T}\psi=0$ then $T\psi=0$. Conversely,
if $T\psi=0$ then $\|\sqrt{T}\psi\|^2 = \ip{\psi}{T\psi}=0$, so $\sqrt{T}\psi=0$.} which in our case implies that $\ker A_+$ contains $\ker A$
and hence every vector of the form $EPf=0$ for test function $f$. Hence 
\begin{equation}
W_\SJ(f,Pg) = \ip{\Gamma f}{A_+Pg}=0, \qquad W_\SJ(Pf,g) = \ip{A_+ P\Gamma f}{g}=0 
\end{equation}
using self-adjointness of $A_+$ and the fact that $P$ commutes with complex conjugation. Therefore $W_\SJ$ is a bi-solution to the Klein--Gordon equation.
Furthermore, $W_\SJ$ has the hermiticity property
\begin{equation}
\overline{W_\SJ(f,g)} = \ip{A_+ g}{\Gamma f} = W_\SJ(\overline{g},\overline{f}).
\end{equation}
The remaining issue is to check the antisymmetric part of $W_\SJ$.
Noting that $A$ is anti-invariant under conjugation with $\Gamma$, but $A^2$ is (consequently) invariant, we have
\begin{equation}
\Gamma A_\pm \Gamma = -A_\mp
\end{equation}
and hence $A=A_+ - \Gamma A_+ \Gamma$. Thus
\begin{align}
iE(f,g) &= W_\SJ(f,g) - \ip{\Gamma f}{\Gamma A_+ \Gamma g} = W_\SJ(f,g) - \ip{A_+\Gamma g}{f}  \\
&=W_\SJ(f,g)-W_\SJ(g,f), 
\end{align}
as required.

Modulo the functional analytic question of whether $A$ is well-defined as a self-adjoint operator on $\HH$, we have now shown that $W_\SJ$ meets all the criteria to define a quasifree state of the real scalar field in either its 
Weyl algebra or infinitesimal Weyl algebra form. It is worth emphasising again that no structures have been introduced here beyond those always available on a globally hyperbolic spacetime, specifically, the volume measure and the advanced-minus-retarded solution operator. As Sorkin has noted,\cite{SorkinGF:2017} only the advanced (or retarded) Green function is required, because the other can be obtained by taking adjoints. 

The analytic questions were resolved by Verch and the present author in Ref.~\refcite{FewsterVerch-SJ}, where we established the viability of the SJ proposal in various circumstances:
 \begin{theorem} \label{thm:SJviable}
 	(a) If $A$ is self-adjoint, then
 	the SJ prescription yields a pure, quasifree state which
 	has distributional $n$-point functions. 
 	(b) In particular, if $\Mb$ can be suitably embedded as a relatively
	compact subset of a globally hyperbolic spacetime then the 
	operator $A=iE$ is a bounded self-adjoint operator on $\HH$.  
\end{theorem}  
Here `suitably embedded' means an embedding of the sort studied in Section~\ref{sec:nogo}, i.e.,  a smooth isometric embedding preserving (time)-orientation and with causally convex image. Note that Theorem~\ref{thm:SJviable} does not address whether the state is Hadamard or is physically reasonable on other grounds.

How does the SJ state get around the no-go theorem discussed earlier?
There are several points to make. At the technical level, Theorem~\ref{thm:SJviable}
does not establish the existence of SJ state in every globally hyperbolic spacetime,
nor that it has the Reeh--Schlieder property in some spacetime. More fundamentally, 
however, there is no reason to expect that the SJ state is locally determined by the spacetime geometry; rather, it is globally determined by the spectral theory of $E_\op$. As discussed earlier,
this raises questions about the operational status of the SJ state, because it depends on the entire future and past history of the universe.
Let us now turn to the question of whether or not SJ states are Hadamard.

\subsection{A computation on ultrastatic slabs}\label{sec:ultrastatic}

It will help to have explicit examples of SJ states for 
\emph{ultrastatic slab} spacetimes, i.e.\ spacetimes given as manifolds by
$(-\tau,\tau)\times\Sigma$ with metric $g(\textrm{d}x,\textrm{d}x)= \textrm{d}t^2-h_{ij}\textrm{d}x^i\textrm{d}x^j$, where $(\Sigma,h)$ is any compact Riemannian manifold. For simplicity we restrict to the minimally coupled Klein--Gordon equation with nonzero mass,
\begin{equation} 
\left(\frac{\partial^2}{\partial t^2} +K\right)\phi = 0, 
\qquad  K = -\triangle_h + m^2,
\end{equation}
where $\triangle_h$ is the Laplace--Beltrami operator. It is standard that 
$L^2(\Sigma)$ has a complete basis of $K$-eigenfunctions $K\psi_j=\omega_j^2 \psi_j$,
where the $\omega_j$ may be assumed strictly positive and arranged in non-decreasing order, labelled by a countable set $J$. As $K$ commutes with complex conjugation, 
the basis may be chosen so that each
$\overline{\psi_j}$ also belongs to the basis. In this way, there is
an involution $j\mapsto \bar{j}$ on $J$ so that $\overline{\psi_j} =\psi_{\bar{j}}$.
A standard fact is that the integral kernel of $E$ is
\begin{equation}
E(t,\ux;t',\ux') = \sum_{j\in J} \frac{\sin\omega_j (t'-t)}{\omega_j} \psi_j(\ux)\overline{\psi_j(\ux')},
\end{equation}
from which $A=iE_\op$ is easily seen to be a direct sum of rank-$2$ operators
in the subspaces of $\HH$ spanned by $\psi_j(\ux)e^{\pm i\omega_j t}$.

Extracting the positive part $A_+$ of $A$ reduces to solving a family of $2\times 2$ matrix problems, leading to the SJ two-point function 
\begin{equation}\label{eq:SJ2pt}
W_{SJ}(x;x') = \sum_{j\in J} 
\frac{\Nc_j}{2\omega_j}
(
e^{-i\omega_j t} + i \delta_j \sin \omega_j t )
(
e^{i\omega_j t'} - i \delta_j \sin \omega_j t') \psi_j(\ux)\overline{\psi_j(\ux')},
\end{equation}
where
\begin{equation}\label{eq:deltaj}
\delta_j= 1-\Nc_j^{-1} 
= 1 - \sqrt{1 + \frac{2\sinc 2\omega_j \tau}{1-\sinc 2\omega_j \tau}}
= -\sinc 2\omega_j \tau + O((\omega_j\tau)^{-2})
\end{equation}
and convergence is understood in the sense of distributions, i.e.\ the series is summed after the summands have been integrated against test functions.  

The dependence of $\delta_j$ on $\tau$ shows that $W_\SJ$ depends nonlocally
on the spacetime geometry: the restriction of the SJ two-point function
to a slab with a smaller value of $\tau$ does not give the SJ two-point function for the smaller slab.

Further insight is given by comparing $W_\SJ$ with the two-point function
\[
W_0 (x;x') = \sum_{j\in J} \frac{e^{-i\omega_j (t-t')}}{2\omega_j}\psi_j(\ux)\overline{\psi_j(\ux')},
\]
of the ultrastatic ground state, which is known to be Hadamard.\cite{FullingNarcowichWald} 
In the limit $\tau\to\infty$, one finds that 
$W_{SJ}\to W_{0}$ as distributions, showing that the SJ proposal gives a natural answer in this
asymptotic limit (though note that it is unclear whether one can directly define an SJ state for the full ultrastatic spacetime). 

Our main interest, however, is in the difference $W_\SJ-W_0$ for finite $\tau$, because the
SJ state is Hadamard if and only if the difference is smooth. In Ref.~\refcite{FewsterVerch-SJ}, we argued as follows. If the SJ state is Hadamard, then any derivative of $W_\SJ-W_0$ must be continuous on 
$M\times M$ and hence square-integrable on $M'\times M'$ where $M'=(-\tau',\tau')\times\Sigma$, with $\tau'<\tau$, is a slightly smaller slab spacetime.
In particular, any derivative of $W_\SJ-W_0$ defines the integral kernel of a (self-adjoint) Hilbert-Schmidt operator on $L^2(M')$, which must therefore have a countable set of nonzero eigenvalues 
that are square-summable, and in particular accumulate only at zero. The eigenvalues can again be read off from a family of $2\times 2$ matrix problems, and can be bounded in terms of $\omega_j$ and $\delta_j$ defined above. This reasoning leads to the conclusion that $W_\SJ-W_H$ is
twice continuously differentiable on $M\times M$ only if 
\begin{equation}
\sin 2\omega_j \tau\to 0
\end{equation}
as $j\to\infty$. Accordingly, the $\omega_j$ must become ever closer to integer multiples of $\pi/(2\tau)$ as $\omega_j\to\infty$, which is a very unstable condition with respect to changes of $\tau$
(recall that the $\omega_j$ are fixed by the choice of the spatial section). Indeed, we proved the following: 
\begin{theorem}
	On an  ultrastatic slab $M=(-\tau,\tau)\times \Sigma$,
	the set of $\tau$ for which the SJ state is Hadamard is at most a set of measure zero. Moreover, if $(\Sigma,h)$ is either a flat $3$-torus or round $3$-sphere, then the SJ state is not Hadamard for any  $\tau$.  
\end{theorem} 
 
The SJ state has also been investigated in other spacetime regions of interest, including 
diamond regions in $1+1$-dimensional Minkowski spacetime\cite{Afshordietal:2012} and a patch of the `pair of trousers' $1+1$-dimensional topology-changing spacetime,\cite{BuckDowkerJubbSorkin:2017} studying the massless scalar field in each case.
For the diamond, it was found that the SJ two-point function approaches the Minkowski form when the two points are both near the centre of the diamond and close to one another, relative to
the overall length scale of the diamond. The calculation was a mix of exact summation and numerics and indicates that the SJ state two-point function differs only slightly from the Minkowski vacuum two-point function in this regime. What seems not to be clear is whether the deviation is smooth, or only has finite differentiability, which determines whether the state is Hadamard; it would be interesting to resolve this question. At the corners, the SJ two-point function approaches the two-point function of the ground state on a half-space with reflecting boundary conditions.
While the question of whether the SJ two-point function is Hadamard in the interior of a diamond in is not resolved (certainly, not in general dimensions or for massive fields), it is known that
the SJ state of a Minkowski diamond cannot be extended as a Hadamard state beyond the diamond, by general results in Ref.~\refcite{FewsterVerch-NecHad}. In other words, some singular behaviour
must be present at the boundary, consistent with the results of Ref.~\refcite{Afshordietal:2012}.
The trousers case is interesting because it not globally hyperbolic; here, specific non-Hadamard behaviour (divergent energy density) is found, which is attributed to singular behaviour where the metric degenerates.\cite{BuckDowkerJubbSorkin:2017}

\subsection{The importance of being Hadamard}

The objection that SJ states generically fail to be Hadamard is only convincing if one has accepted that Hadamard states are the largest class of physically acceptable states. As formulated in Section~\ref{sec:Hadamard}, the Hadamard condition 
is a constraint on behaviour at ultra-high energies or ultra-short distances. It is  therefore not a compelling condition if one views continuum QFT as an approximation to a theory on a discrete structure, or formulates it with an energy cut-off.  For example, 
Buck et al remark that outside a full understanding of (semiclassical) quantum gravity,
\emph{Hadamard behavior seems irrelevant ``operationally'', since it corresponds in the Wightman function to the absence of a term that could only be noticed at extremely high energies.}\cite{Afshordietal:2012}
Prompted by such concerns (and following a comment by Klaus Fredenhagen), the author and Verch addressed the fundamental question of how the Hadamard condition can be motivated without explicit reference to these limits or
to point-splitting prescriptions.\cite{FewsterVerch-NecHad} 

The ultrastatic slab with compact spatial sections again provides a useful test-bed, because we can easily construct and study a wide class
of states that encompasses essentially all pure quasifree states associated with the $(t,\ux)$ separation
of variables. In particular, the ground state, finite temperature states and SJ states are incorporated. Consider the bosonic Fock space $\FF$ over $\ell^2(\NN)$, equipped with countable sets of annihilation and creation operators $a_j$, $a_j^*$ labelled by $j\in\NN$. Given any orthonormal basis $\psi_j$
of $K$-eigenfunctions for $L^2(\Sigma)$ so that $\overline{\psi_j}=\psi_{\bar{j}}$ for some involution $j\mapsto\bar{j}$ on $\NN$, and any set of coefficients $\beta_j\in\CC$ such that $\beta_j=\beta_{\bar{j}}$ and satisfying a bound  $|\beta_j|\le P(\omega_j)$ for some polynomial $P$, we may define a field operator on $\FF$ by 
\begin{equation}
\phi(t,\ux) = \sum_j\frac{1}{\sqrt{2\omega_j}}\left( (\alpha_j e^{-i\omega_jt} + \beta_j e^{i\omega_jt}) \psi_j(\ux) a_j + \text{h.c.}\right),
\end{equation}
where $\alpha_j=\sqrt{1+|\beta_j|^2}$. Up to unitary equivalence, this is essentially the most general Fock space quantization of the field with mode functions taking the form $T(t)X(\ux)$. 
The Fock vacuum vector $\Omega$, annihilated by all the $a_j$, determines a state of the theory, with two-point function 
\begin{equation}\label{eq:gen2pt}
W(t,\ux;t',\ux') = \sum_j \frac{1}{2\omega_j}\left(\alpha_j e^{-i\omega_j t} + \beta_j e^{i\omega_j t}\right) \left(\overline{\alpha_j} e^{-i\omega_j t} + \overline{\beta_j} e^{i\omega_j t}\right). 
\end{equation}
One may easily check that the conditions on the $\alpha_j$ and $\beta_j$ imply that 
$W(x,x')-W(x',x)=iE(x,x')$ as required by the canonical commutation relations. 
 
In particular, the usual ultrastatic ground state corresponds to $\alpha_j=1$, $\beta_j=0$ for all $j$, and appropriate values for the SJ state are easily read off from \eqref{eq:SJ2pt}. 
The question of whether $\Omega$ represents a Hadamard state could be addressed by subtracting the ultrastatic two-point function from \eqref{eq:gen2pt} and testing for smoothness. However the microlocal criterion
\eqref{eq:uSC2} provides a much simpler test. Let $f\in\CoinX{-\tau,\tau}$ and $g\in C^\infty(\Sigma)$. Then the smeared field 
\begin{equation}
\phi(f\otimes g)=\int \phi(t,\ux)f(t)g(\ux) \,dt\,dvol_\Sigma
\end{equation}
is easily seen to obey
\begin{align}
\|\phi(f\otimes g)\Omega\|^2 &= \sum_{j\in J} \frac{1}{2\omega_j}|\alpha_j \hat{f}(\omega_j) + \overline{\beta_j}\hat{f}(-\omega_j)|^2 |\ip{\psi_j}{g}|^2 \nonumber \\
&\le \|g\|^2 \sum_{j\in J} \frac{1}{2\omega_j}|\alpha_j \hat{f}(\omega_j) + \overline{\beta_j}\hat{f}(-\omega_j)|^2 ,
\end{align}
where we have used the Cauchy--Schwarz inequality and $\|g\|$ is the $L^2(\Sigma)$ norm.
For simplicity, suppose that one can choose global coordinates $x^i$ on $\Sigma$ and define
$e_{\zeta,\uk}(t,\ux)=e^{i(\zeta t+\uk\cdot \ux)}$. Then
\begin{equation}
\|\phi([f\otimes g]e_{\zeta,\uk})\Omega\|^2 \le \|g\|^2 \sum_{j\in J} \frac{1}{2\omega_j}|\alpha_j \hat{f}(\omega_j+\zeta) + \overline{\beta_j}\hat{f}(\zeta-\omega_j)|^2. 
\end{equation}
The important point is that increasing $\zeta>0$ pushes $\hat{f}(\omega_j+\zeta)$ further into its tail; on the other hand, $\hat{f}(\zeta-\omega_j)$ has its peak where $\omega_j\approx \zeta$, so there is a risk that the overall contribution from these terms might not decay, or might even grow, as $\zeta\to\infty$. However, if we assume that $\omega_j^N\beta_j \to 0$ as $\omega_j\to\infty$, for every $N\in\NN$ (which also implies that $\alpha_j\to 1$ as $\omega_j\to\infty$) then the right-hand side decays faster than any inverse power of $\zeta$, for $\zeta\to +\infty$. In turn, this implies that every point $(x,k)\in T^*M$ with $k_0>0$ is a regular direction for the distribution $\phi(\cdot)\Omega$. It follows that all
$(x,k)\in\WF(\phi(\cdot)\Omega)$ must have $k_0\le 0$. But $\phi$ is a distributional solution to the Klein--Gordon equation, so its wavefront set is contained in the characteristic set of $P$, which consists of null covectors. Hence
\begin{equation}
\WF(\phi(\cdot)\Omega)\subset\{(x,k): g^{ab}k_a k_b=0~k_0\le 0\} = \Nc^-,
\end{equation}
thus establishing the microlocal spectrum condition. Summarising so far, provided the $\beta_j$ decay sufficiently rapidly, then the state $\Omega$ is Hadamard. 

Our aim in this section is to motivate the Hadamard condition without direct reference to ultra-high energies. So now consider a general sequence of coefficients $\beta_j=\overline{\beta_{\bar{j}}}$, assuming only that they do not grow faster than polynomially in $\omega_j$. As the fields have been constructed 
explicitly in a Fock space, we can construct Wick squares by normal ordering 
the annihilation and creation operators, avoiding the use of point-splitting and Hadamard regulation (or any other prescription based on coincidence limits).
The resulting operators necessarily have vanishing expectation values in the state $\Omega$ but can nonetheless exhibit nonzero \emph{fluctuations}.
For example, the operator 
${:}(\partial_t^k\phi)^2{:}(f\otimes 1)$ has squared fluctuation
\begin{equation}\label{eq:fluct}
\|{:}(\partial_t^k\phi)^2{:}(f\otimes 1)\Omega\|^2 = \frac{1}{2}\sum_{j\in J} \omega_j^{4k-2} 
|(\alpha_j^2+\overline{\beta_j}^2)\hat{f}(2\omega_j) +  (-1)^{k} 2\alpha_j\overline{\beta_j} \hat{f}(0)|^2
\end{equation}
in the state $\Omega$, for any real-valued even test function $f$ supported
in $(-\tau,\tau)$.
These fluctuations are finite for all $k$ only if the $\beta_j$ decay
rapidly as $\omega_j\to\infty$. To see why, note that the rapid decrease of the transform $\hat{f}(\omega_j)$, and our assumption that the $\beta_j$ do not grow faster than polynomially
in $\omega_j$, imply that the convergence of \eqref{eq:fluct}
is determined by whether 
\begin{equation}
\sum_{j\in J} \omega_j^{4k-2} |\alpha_j \overline{\beta_j}|^2
\end{equation}
converges for all $k$. Convergence requires that the summands must tend to zero, 
so, as $\alpha_j\ge 1$, it follows that $\beta_j$ must tend to zero faster than any inverse power in $\omega_j$. Consequently, by our earlier arguments, the state $\Omega$ must be Hadamard. In this way, the Hadamard condition can be motivated without reference to short-distance or high-energy behaviour, but simply by the requirement 
that all normal-ordered Wick squares of derivatives of the field should have finite fluctuations about their (vanishing) expectation values.

This argument provides a partial converse to results of Brunetti, Fredenhagen and K\"ohler,\cite{BruFreKoe} which imply that Wick polynomials
defined relative to a Hadamard reference state have
finite fluctuations in that state.  In summary, we have established:
\begin{theorem}
	A state of the scalar field on the ultrastatic slab with two-point function \eqref{eq:gen2pt} defines a normal-ordering in which all Wick squares of field derivatives have finite fluctuations if and only if it is Hadamard. 
	Equivalently, the coefficients $\beta_j$ tend to zero faster than any inverse power of $\omega_j$ as $\omega_j\to\infty$.
\end{theorem}
It is an open problem to extend this result to general spacetimes or more general
states of the field on the ultrastatic slab (recall that our family of states is closely related to separation of variables in the $t,\ux$ coordinates).

Turning our result around, any attempt to use a non-Hadamard state as a foundation
for quantum field theory will be limited in the range of Wick polynomials available.
Indeed, perturbation theory would simply be impossible beyond a finite order.\footnote{I thank Daniel Siemssen for discussions on this point.} Finiteness can be restored by imposing an energy cut-off, but at the price that results diverge as the cut-off is removed (even after regulation by normal ordering). Therefore at large finite values of the cut-off, the fluctuations would typically be unphysically large. 

\subsection{BF states}

The previous subsection makes clear that SJ states on the slab are not Hadamard because the relevant coefficients $\beta_j$ do not decay sufficiently fast.   A clever modification of the SJ prescription, 
due to Brum and Fredenhagen,\cite{BrumFredenhagen:2014} resolves this problem and provides a class
of Hadamard states at least in some cases.
 
As in our Theorem~\ref{thm:SJviable}, suppose $\Mb$ is
isometrically embedded as a relatively compact subset of $\Nb$. 
Now choose any $\chi\in\CoinX{\Nb}$ with $\chi\equiv 1$ on $\Mb$ and
define $A_{\textrm{BF}} = i \chi E_\Nb \chi$ as a symmetric operator
on $L^2(\Nb)$. Taking the positive part $A_{\textrm{BF}+}$ of this operator,
the BF two-point function is
\begin{equation}
W_{\textrm{BF}}(f,g)=\ip{\overline{f}}{A_{\textrm{BF}+}g} \qquad(f,g\in\CoinX{\Mb}).
\end{equation}
On restriction to $\Mb$, $W_{\textrm{BF}}$ defines a quasifree state of the scalar field. In the ultrastatic slab example, the effect of this modification is to replace the $\sinc$ functions in the formula \eqref{eq:deltaj} by the Fourier transform of $\chi^2$ (up to factors). Consequently, $\delta_j$ and $\beta_j = \frac{1}{2}\delta_j (1-\delta_j)^{-1/2}$ decay faster than any inverse power of $\omega_j$. Consequently, the resulting \emph{BF state} is Hadamard. Brum and Fredenhagen actually proved this in a more general setting than the
ultrastatic slab -- they considered static or cosmological spacetimes of compact spatial section. 
 
The key difference between the SJ and BF prescriptions, in that the former aims to construct a single distinguished state, while the latter produces
a family of Hadamard states parametrised by the function $\chi$ and the embedding
of $\Mb$ into a larger spacetime. As explicit constructions of Hadamard states are
few and far between, this is an important contribution.

\subsection{FP states for Dirac fields}

We have already mentioned that the SJ construction shares similarities with Finster's fermionic projector construction, which arose from attempts to
define a Dirac sea in curved spacetimes. The parallel can be seen clearly in
Ref.~\refcite{FinsterReintjes:2015}, for example, and the corresponding `FP states' were constructed and investigated by Lang and the author in Ref.~\refcite{FewsterLang:2015}.

Consider a Lorentzian spin manifold $\Mb$. Every
solution to the Dirac equation with spacelike compact support may be written
$\psi= Su$ where $u$ is a test spinor field and $S$ is the difference
of advanced and retarded Green operators. These solutions
form a pre-Hilbert space with respect to the inner product
\begin{equation}\label{eq:Dirac_ip}
\ip{Su}{Sv} = i \int_\Mb u^+ S v \,\dvol_\Mb
\end{equation}
and we may complete to obtain a Hilbert space $\HH$. Here $u^+=u^\dagger \gamma^0$ is the Dirac adjoint. For definiteness,
we work on ultrastatic slab spacetimes with compact Cauchy surfaces.
The slab can be embedded in the full ultrastatic spacetime 
$\Nb$ and any solution $\psi$ on $\Mb$ can be extended to a solution 
$\tilde{\psi}$ on $\Nb$. Given any integrable function $f$ on $\Nb$, 
we may define a bounded self-adjoint operator on $\HH$ by
\begin{equation}
\ip{\psi}{A_f\varphi} = \int f \tilde{\psi}^+ \tilde{\varphi}\, \dvol_\Nb
\end{equation}
and the spectral projection of $A$ onto the positive half-line $P_{\RR^+}(A_f)$
can be used to define a pure quasi-free state $\omega_{\textrm{FP}}$ of the Dirac field with
\begin{equation}
\omega_{\textrm{FP}}(\Psi(v)\Psi^+(u)) = \ip{Sv^+}{P_{\RR^+}(A_f) Su},
\end{equation}
where $\Psi(v)$ and $\Psi^+(u)$
are the Dirac spinor and cospinor fields, smeared respectively with cospinor or spinor test sections. This state is pure, quasifree and gauge-invariant. 

The simplest choice for $f$ is the characteristic function of $\Mb$,
and is also the closest in spirit to the fermionic projector construction in 
Ref.~\refcite{FinsterReintjes:2015}.
The resulting \emph{unsoftened FP state} is the direct analogue of the SJ state, and generally fails to be Hadamard. However, if $f$ is nonnegative, smooth and compactly supported, then $\omega_f$ is a \emph{softened FP state} (analogous to a BF state) and is Hadamard.  

As in the scalar case, one can investigate the link between finite fluctuations
of Wick polynomials and the Hadamard condition. Among FP states, it turns out that
finite fluctuations imply the Hadamard condition and vice versa.\cite{FewsterLang:2015} More generally,
however, we conjecture that a state of the Dirac field exhibiting finite fluctuations for all Wick polynomials is either Hadamard or \emph{anti-Hadamard}.  Here, an anti-Hadamard state is one whose microlocal properties are precisely reversed relative to those of Hadamard states (there are no states of this type for the scalar field).  
Finally, we mention that the construction can be extended to suitable spacetimes of
arbitrary lifetime by adopting a `mass oscillation' principle, in which one
integrates over the mass parameter in a family of solutions, before taking the spacetime integral in~\eqref{eq:Dirac_ip}, and which also gives a Hadamard state in certain circumstances.\cite{FinsterReintjes:2016,FinsterMurroRoeken:2016}

\section{Summary} 

The overarching message of this paper is that attempts to find
a prescription for determining a single state on general spacetimes
will always encounter problems: if the prescription is local then the states
produced cannot have good physical properties; if it is nonlocal then the operational
status of the states is unclear, because it is not clear how a local
experimenter would prepare it even approximately. 

In particular, we have seen that the SJ proposal, in its original formulation, 
is nonlocal and produces states that can fail to be Hadamard. Nonetheless,
the SJ proposal has provoked several useful developments: both in terms
of novel concrete constructions of classes of Hadamard states by softening the SJ construction, and by focussing attention on how to motivate the Hadamard condition
without explicit reference to short-distance structure. 

A number of directions are open: first, one would like a general proof that states with finite fluctuations for Wick polynomials are necessarily Hadamard (or possibly anti-Hadamard, in the Dirac case). Second, the detailed physical properties of SJ, BF, and FP states remain far from clear. 
For example, while it is known that SJ states fail to be Hadamard on slab spacetimes, 
the corresponding question is open for diamond regions in Minkowski spacetime (as discussed
at the end of subsection~\ref{sec:ultrastatic}). Regarding BF and FP states, one would like to 
prove that they are Hadamard under more general circumstances. Third, within the BF and FP classes of Hadamard states, can the softening function be tuned so as to 
produce recognizably vacuum-like behaviour in large subregions of spacetime? Finally, we mention that it is possible to extend similar constructions to the Proca field. Here, however, one has to contend
with additional technical complications because the natural inner product space of
vector potentials is indefinite. Furthermore, the analogue of the advanced-minus-retarded operator is unbounded. Surprisingly, it turns out that 
suitably softened BF-type states exist and are Hadamard, while SJ-type states fail to exist. These results will be reported on elsewhere, in joint work with Rejzner and Wingham. 
 
\section*{Acknowledgments}

I thank Kasia Rejzner and Daniel Siemssen for comments on the manuscript.

 \end{document}